\documentclass[final,leqno,onefignum,onetabnum]{siamltex1213}

\usepackage{graphicx}
\usepackage{epstopdf}
\input{preamble.tex}

\usepackage{enumitem}

\graphicspath{{.}{./fig/}}

\title{AptRank: An Adaptive PageRank Model for Protein Function Prediction on Bi-relational Graphs}

\author{Biaobin Jiang\footnotemark[1] \and Kyle Kloster\footnotemark[2] \and
David F. Gleich\footnotemark[3] \and Michael Gribskov\footnotemark[1]\footnotemark[3]}

\renewcommand{\abstractname}{}

\begin{document}
\maketitle%
%\slugger{simax}{xxxx}{xx}{x}{x--x}

%\newcommand{\citep}{\cite}
%\newcommand{\RR}{\mathbb{R}}
%\newcommand{\colst}[1]{\mathcal{C}\left(#1\right) }

\renewcommand{\thefootnote}{\fnsymbol{footnote}}

\footnotetext[1]{Department of Biological Sciences, Purdue University, West Lafayette, IN. }
\footnotetext[2]{Department of Mathematics, Purdue University, West Lafayette, IN.}
\footnotetext[3]{Computer Science Department, Purdue University, West Lafayette, IN.}

\renewcommand{\thefootnote}{\arabic{footnote}}

\begin{abstract}

\noindent \textbf{Motivation.} Diffusion-based network models are widely used for protein function prediction using protein network data and have been shown to outperform neighborhood-based and module-based methods. Recent studies have shown that integrating the hierarchical structure of the Gene Ontology (GO) data dramatically improves prediction accuracy. However, previous methods usually either used the GO hierarchy to refine the prediction results of multiple classifiers, or flattened the hierarchy into a function-function similarity kernel. No study has taken the GO hierarchy into account together with the protein network as a two-layer network model. \medskip

\noindent \textbf{Results.} We first construct a Bi-relational graph (Birg) model comprised of both protein-protein association and function-function hierarchical networks. We then propose two diffusion-based methods, BirgRank and AptRank, both of which use PageRank to diffuse information on this two-layer graph model. BirgRank is a direct application of traditional PageRank with fixed decay parameters. In contrast, AptRank utilizes an adaptive diffusion mechanism to improve the performance of BirgRank. We evaluate the ability of both methods to predict protein function on yeast, fly, and human protein datasets, and compare with four previous methods: GeneMANIA, TMC, ProteinRank and clusDCA. We design three different validation strategies: missing function prediction, \textit{de novo} function prediction, and guided function prediction to comprehensively evaluate predictability of all six methods. We find that both BirgRank and AptRank outperform the previous methods, especially in missing function prediction when using only $10\%$ of  the data for training. \medskip

\noindent \textbf{Conclusion.} AptRank naturally combines protein-protein associations and the GO function-function hierarchy into a two-layer network model without flattening the hierarchy into a similarity kernel. Introducing an adaptive mechanism to the traditional, fixed-parameter model of PageRank greatly improves the accuracy of protein function prediction.\medskip

\noindent \textbf{Code.} \url{https://github.rcac.purdue.edu/mgribsko/aptrank}. \\
\textbf{Contact.} \href{gribskov@purdue.edu}{gribskov@purdue.edu}
\end{abstract}

\pagestyle{myheadings}
\thispagestyle{plain}
\markboth{Jiang \textit{et~al}.}{An Adaptive PageRank Model for Protein Function Prediction}

\section{Introduction}

Given a set of functionally uncharacterized genes or proteins from a Genome-Wide Association Study, or differential expression analysis,
experimental biologists often have little \emph{a priori} information available to guide the design of hypothesis-based experiments to determine molecular functions. For example, what is the expected phenotype if a particular gene is removed? It would greatly improve hypothesis formation if biologists had prior insight from predicted functions of interesting genes or proteins in databases. Computational annotation of genes or proteins with unknown functions is thus a fundamental research area in computational biology.

In the past decade, there has been much work to accurately predict functional annotations of genes or proteins using heterogeneous molecular feature data \citep{critical2008mouse,eval2013dk}.
The collected molecular features include gene expression, sequence patterns, evolutionary conservation profiles, protein structures and domains, protein-protein interactions (PPIs), and phenotypes or disease associations.
In one comprehensive assessment \citep{critical2008mouse}, one of the methods, GeneMANIA \citep{genemania} slightly outperformed the other eight methods by integrating the multiple molecular features into a functional association network (a.k.a., a kernel).
The success story of GeneMANIA suggests two important ideas.
First, we can significantly improve prediction methods that rely on a single data type by integrating data of many types. And second, kernel integration is a particularly powerful approach to combining multiple types of data.

Given an integrated functional association network, methods for protein function prediction can be divided into three different types: neighborhood-based, module-assisted, and diffusion-based \citep{sharan2007network}.
Neighborhood-based methods \citep{schwikowski2000network} predict the function of one protein by using the functions of its neighbors in the network, i.e., the guilt-by-association approach. This approach has two obvious drawbacks. On one hand, it ignores the functional information from all the other proteins outside the neighborhoods of the query proteins, which leads to a low true-positive rate. On the other hand, it may also have high false-positive rates when the query protein has a single function but is surrounded by many multi-functional proteins.

Module-assisted methods operate by first partitioning a network or a kernel into functional modules \citep{enright2002efficient,bader2003automated}. Biologically, a functional module in a PPI network is a group of physically interacting proteins engaged in a biological activity, e.g., to form a scaffold or to relay signals. In network science, a good module is commonly defined as a densely connected subgraph with loose connections to the outside \citep{newman2004finding}. This definition is naturally coincident with protein complexes, but not signaling cascades. Obtaining a high-quality graph partition is challenging, and this field of study is still highly active.

Diffusion-based methods generally simulate propagating information from functionally known proteins to unknown ones through network connectivity. \citet{nabieva2005whole} constructed a network flow model with fixed diffusion distances and capacities on network edges. This method was claimed to capture both global network topology as well as local network structure to improve the function predictability over the first two domains of methods mentioned above. \citet{freschi2007protein} devised a tool called ProteinRank by utilizing PageRank \citep{page1999pagerank}, the method used by Google to rank webpages, to diffuse functional annotation information throughout a network without setting a fixed diffusion distance or edge capacities. \citet{genemania} utilized the Label Propagation algorithm \citep{zhou2004learning} to develop GeneMANIA as a classification model with multiple heterogeneous network datasets using weighted kernels and labeled negative samples. The method achieved approximately $70\sim 90\%$ accuracy in three-fold cross validation using a benchmark dataset \citep{critical2008mouse}. \citet{yu2013protein} developed the Transductive Multilabel Classifier (TMC), based on a Bi-relational graph \citep{wang2011image} consisting of a protein interactome and cosine similarities in a protein functional profile as two kernels in each graph layer. Then they used PageRank on this two-layer graph to diffuse functional information to predict protein functions.

Functional annotation data are usually organized in a \emph{tree-like} ontological structure with general terms at the root and specific terms on the leaves \citep{go2004gene}. However, the majority of previous methods disregard this intrinsic hierarchical structure by assuming that the relationships between functions are independent.
Recently, several methods have been proposed in order to take into account the interdependent relationships between functional terms in the hierarchical structure.
\citet{king2003} predicted gene functions using decision trees and Bayesian networks while taking advantage of the annotation dependency between different branches of the GO hierarchy. %\citep{king2003}.
Notably, when they trained and tested the association of functional terms with genes, they excluded the information from any ancestors and descendants of the terms in question. This ensures a fair cross validation in which prediction does not benefit from the GO annotation rule: if one gene is annotated by a term, then that gene is automatically annotated by all the ancestors of that term.
\citet{barutcuoglu2006hierarchical} and \citet{valentini2011true} proposed a hierarchical Bayesian framework and a True Path Rule, respectively, to perform ensemble learning of the classification results yielded by multiple Support Vector Machines (SVMs). They demonstrated that the accuracy of protein function prediction can be significantly improved by integrating the functional hierarchy \citep{valentini2014hierarchical}. \citet{tao2007information} and \citet{pandey2009incorporating} utilized Lin's similarity \citep{lin1998information} to flatten the functional hierarchy, and then predicted protein functions using a \emph{k-}Nearest Neighbor (\emph{k}-NN) method. \citet{gostruct} directly modeled the hierarchical structure of functional ontology using structured SVM \citep{structsvm}, and showed that their method outperformed \emph{k}-NN and other binary classifiers without taking the hierarchy into account. Recently, \citet{yu2015down} combined Lin's similarity of protein functional profiles with an ontological hierarchy using downward random walks with restarts, so as to improve the TMC model \citep{yu2013protein}, which can predict functions of a protein that are not in its neighborhood, but are present in the hierarchy. \citet{wang2015exploiting} proposed clusDCA for protein function prediction by integrating protein networks and a functional hierarchy, using PageRank for network smoothing and low-rank matrix approximation to de-noise the network data.

In this study, we propose two methods that directly diffusing information on the functional hierarchy other than a flat functional similarity constructed by Lin's method \citep{lin1998information}.
The first method, which we call BirgRank, constructs a Bi-relational graph model with a protein-protein functional association network as one layer and an unflattened ontological hierarchy as a second layer, and then directly applies  PageRank to diffuse annotation information across the two-layer network. The second method, which we call AptRank, employs an adaptive version of PageRank that replaces the standard PageRank parameters with values dynamically chosen to better fit the training data.
The main differences between our methods and other diffusion-based methods are (1) we do not require any negative labeled samples since our method is not a traditional classification model; (2) we take full advantage of the functional hierarchy as a two-way directed graph, and do not use Lin's similarity \citep{lin1998information}, or any kernel trick, to flatten the hierarchy; and (3) we avoid using the annotation of a particular term to predict the annotation of its parental terms, we train and test our methods using the direct annotations only (see Figure~\ref{fig:example}(B) and (C)), which guarantees that the functional terms to be tested for each protein are mutually neither ancestors nor descendants in the GO hierarchy.

To avoid the inflated accuracies of network-based methods in protein function prediction noted by Gillis and Pavlidis \citep{gillis2011impact,gillis2012guilt,pavlidis2013progress,gillis2014bias}, we conduct a large and strict evaluation of our methods against the other state-of-the-art methods. In addition to three small benchmark datasets, we use an up-to-date protein interaction network dataset and exclude the functional annotations inferred from protein interactions (evidence code: IPI). Rather than two-fold \citep{freschi2007protein}, three-fold \citep{genemania,wang2015exploiting} or five-fold \citep{yu2013protein} cross validation, we design three different validations: missing function prediction, \textit{de novo} function prediction, and a hybrid of the two strategies, namely guided function prediction.
For each of the three types of validation, we perform the validation method using $20\%$ or $10\%$ of the data in training.
To overcome the drawback of using Area Under the ROC curve (AUROC) as a criterion in evaluating performance on imbalanced data with a small number of positive samples, we also utilize Mean Average Precision (MAP) which focuses on the ranking of positive samples only, and is widely used in the field of information retrieval.

\section{Methods}

\subsection{Problem Statement}
\begin{figure}[!ht]
        \centering
        \includegraphics[width = 0.618\textwidth]{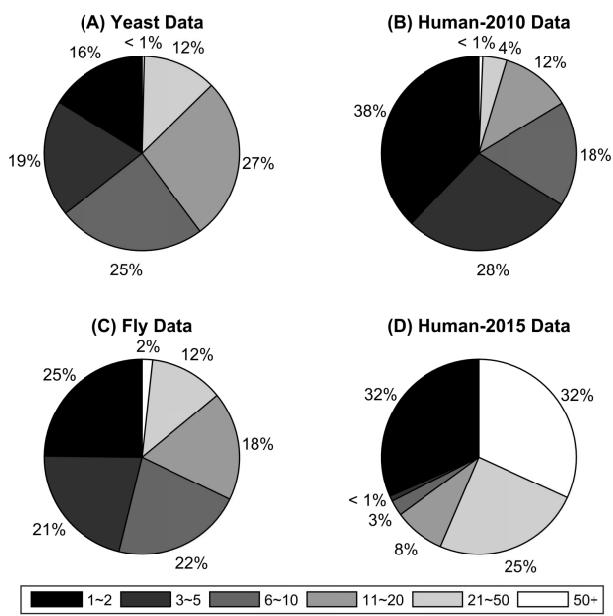}\\
        \caption{\label{fig:distfun} Distribution of annotated functions of proteins in (A) yeast, (B) human collected in 2010, (C) fly and (D) human collected in 2015. The yeast, human-2010, and fly datasets are collected from and maintained by GeneMANIA developers \citep{swdataweb}.}
    \end{figure}

This study is motivated by the fact that there are still many proteins whose functions are poorly characterized. To examine the extent to which each protein has been experimentally annotated, we downloaded three benchmark datasets of yeast, fly and human proteins maintained by GeneMANIA-SW since $2010$ \citep{swdataweb}, and also the human Gene Ontology Annotation (GOA) data \citep{gene2015gene} in March $2015$. For the human GOA data, we only consider the annotations in the Biological Process (BP) category, regardless of Molecular Function (MF) and Cellular Component (CC) terms. Also, we only use annotations with experimental evidence codes, within which we remove the terms inferred by physical interaction (evidence code: IPI). All of these four datasets will be used for evaluation later in this study. We illustrate the proportion of the number of functional annotations of each protein in Figure~\ref{fig:distfun}. We can see that there are a large number of proteins with fewer than $3$ functional annotations. This is primarily due to bias in biological research interests and the difficulty of experimentally determining protein functions.

The aim of this study is to predict protein functions given a protein-protein association network and a hierarchically structured set of functional terms. The hypothesis is that associated proteins in the protein network are likely to share similar functions. Here, we define a protein-protein association network as pairwise quantitative relationships of proteins. This network either can be sparse and binary, e.g., a protein-protein physical interaction network, or weighted and dense, e.g., a pairwise similarity of protein sequences.

\subsection{Preliminaries of Personalized PageRank}
PageRank is a well-studied model in network analysis that simulates how information diffuses across a network~\citep{page1999pagerank}. It is also called Random Walk with Restart (RWR) in other literature \citep{tong2006fast}. We will use PageRank to diffuse annotation information from well-annotated proteins through a functional association network to less well-annotated proteins.
In particular, we use a ``personalized'' variation of PageRank~\citep{jeh2003-personalized}, which models the flow of information from a small number of specific objects, called source nodes (in our case, a single protein) to the remainder of a network. And we use this model to quantify which functions are most relevant to a source protein.

\begin{figure}[!ht]
        \centering
        \includegraphics[width = 0.618\textwidth]{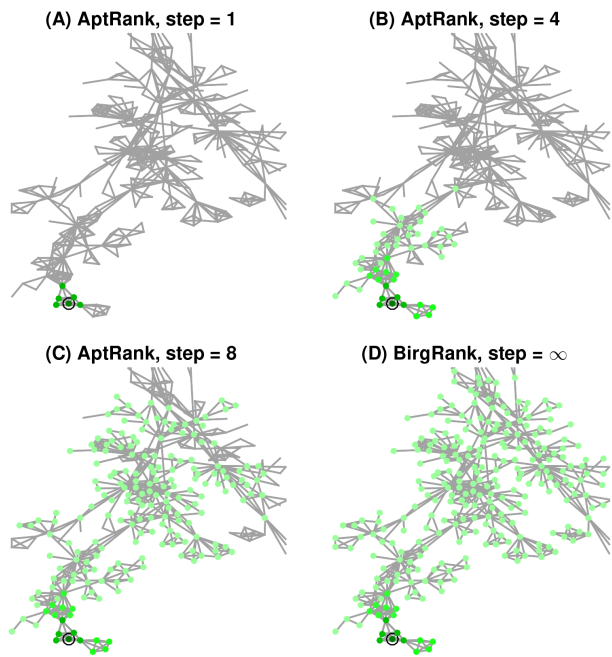}\\
        \caption{\label{fig:diffusion}
        Information diffusion in Personalized PageRank. Diffusion starts from the node circled in black. The green dye diffuses from the black circled node. Nodes where the diffusion concentrates the most appear the darkest green; this indicates the nodes that are most strongly connected to the black circled node. (A), (B) and (C) illustrate our AptRank diffusion with different step sizes. (D) displays our BirgRank diffusion once the associated Markov chain has converged to its stationary distribution.
        }
    \end{figure}

Intuitively, personalized PageRank operates on a network of interconnected nodes by placing a quantity of ``dye" at a source node of interest, then letting the dye diffuse across the edges of the network, decaying as it spreads. Once the diffusion process decays to zero, the network regions where the largest amount of dye has concentrated are then the most important regions to the source node. See Figure~\ref{fig:diffusion} for a visualization of the dye diffusing from a source node.

Mathematically, on a network with $n$ objects, the network is modeled by an adjacency matrix $\mA \in \RR^{n \times n}$ such that $\mA_{ij}$ is 1 if node $j$ has an edge to node $i$, and is 0 otherwise. To model the diffusion process beginning with ``dye" at a source node,
we use a vector $\vv \in \RR^{n \times 1}$ that is all $0$s except for a $1$ in the entry corresponding to the source node. This vector $\vv$ is called the personalization vector. Let $\vx \in \RR^{n \times 1}$ be a vector representing the amount of dye at each node in the network at some point during the diffusion process. We then model the diffusion of the dye across the graph by multiplying $\vx$ by a column-stochastic version of $\mA$; this represents the dye on node $j$ being distributed in equal parts to each neighbor $i$ of node $j$. We denote the column-stochastic version of any nonnegative matrix $\mM$ as $\bar{\mM}$; this is computed by dividing each column of the matrix $\mM$ by the sum of the entries in that column.

Finally, the decay of the diffusion process is controlled by the so-called PageRank teleportation parameter, $\alpha \in (0,1)$.
During each stage of the diffusion, the dye that spreads across the network decays proportionally to $\alpha$, so that the amount of dye still diffusing after $k$ steps is $\alpha^k$.
Then the PageRank vector $\vx$ is given by the solution of the linear system
\begin{equation}\label{eq:ppr}
  (\mI - \alpha \bar{\mA}) \vx = (1 - \alpha)\vv.
\end{equation}
Recall our intuition that the PageRank vector indicates how much of the dye flows from the source node (i.e. the nonzero entry in the vector $\vv$) to each node in the graph. In our context, this means that $\vx$ will indicate how much of the functional information flows from the protein of interest to each other protein in the graph. In our model, we combine proteins and functions into a single network so that the PageRank vector can indicate diffusion flow between proteins and functions.

The solution to the Personalized PageRank linear system in Equation~\eqref{eq:ppr} can be expressed as
\begin{equation}\label{eq:prseries}
  \vx = \sum_{k=0}^{\infty} (1-\alpha)\alpha^k \bar{\mA}^k \vv.
\end{equation}
This expression will become useful when we introduce the idea of using adaptive coefficients in place of $\alpha^k$ to optimize prediction quality (see Section~\ref{sec:aptrank}).
We note that, although PageRank has an interpretation as a Markov chain, and Markov chains must meet certain conditions to guarantee convergence to a stationary distribution, this matrix power series~\eqref{eq:prseries} always converges for any $\alpha \in (0,1)$ and stochastic matrix $\bar{\mA}$. Thus, the existence of the unique solution $\vx$ is guaranteed regardless of the structure of the matrix $\mA$.
We emphasize this because the form of linear system that we use differs from the traditional PageRank setting, which uses Markov chain analysis in the proof of its convergence; in contrast, our computations do not rely on this Markov chain analysis. %Instead, convergence of relying simply on the convergence of the matrix power series.

%%%%%%%%%%%%%%%%%%%%%%%%%%%%%%%%%%%%%%%%%%%%%%%%%%%%%%%%%

\subsection{BirgRank: Bi-relational graph PageRank model}
\begin{figure}[!ht]
        \centering
        \includegraphics[width = 0.618\textwidth]{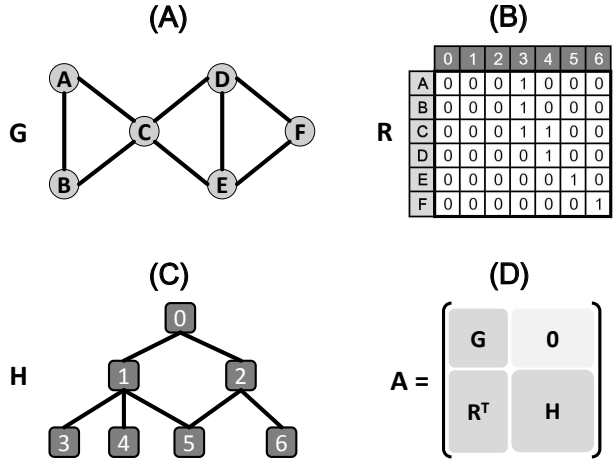}\\
        \caption{\label{fig:example} Given data visualization using simple example. (A) protein-protein binary interaction network, (B) protein-function reference matrix, (C) function-function hierarchy, (D) adjacency matrix $\mA$ of a bi-relational graph.}
    \end{figure}

We denote the number of proteins by $m$ and the number of function terms by $n$. Then the three given datasets (protein-protein association network, protein-function annotations, and function-function hierarchy) are denoted by the following matrices:
\begin{itemize}
  \item $\mG \in \RR^{m \times m}$, a symmetric matrix where $G(i,j)$ denotes to which extent protein $i$ is associated with protein $j$;
  \item $\mR \in \RR^{m \times n}$, a binary matrix where $R(i,j) = 1$ if protein $i$ is annotated by function $j$, 0 otherwise; and
  \item $\mH \in \RR^{n \times n}$, a binary matrix where $H(i,j) = 1$ if functional term $i$ is the child of term $j$, 0 otherwise.
\end{itemize}
We illustrate these three components in Figure~\ref{fig:example}(A), (B) and (C), using a small example with $6$ proteins and $7$ functional terms. For simplicity, Figure~\ref{fig:example}(A) shows a protein-protein binary interaction network, but it can be replaced by any protein-protein association network. Functional terms are hierarchically structured in a Gene Ontology (Figure~\ref{fig:example}(C)) like an upside down ``tree'', where the terms on the top (root) are more general and the ones in the bottom (leaves) are more specific. The annotation rule is that if one gene/protein is annotated by one term, then this gene/protein is automatically annotated by all the parental terms of that term in the hierarchy. However, note that in this study we only consider training and predicting the direct annotations of each protein, and do not propagate the corresponding parental annotations using the annotation rule, as shown in Figure~\ref{fig:example}(B). This ensures that our prediction does not benefit from the annotation rule.

Next, we construct a bi-relational graph \citep{wang2011image} that incorporates these three datasets into a single network (Figure~\ref{fig:example}(D)).
To evaluate prediction performance, we split all the annotations in $\mR$ into $\mR_T$, which we use for training during model construction, and $\mR_E$, which we use for evaluating predictions (see Figure~\ref{fig:split}). For each protein $i$, we predict its functions using Equation~\eqref{eq:ppr} by setting it as the diffusion source, i.e., by computing the diffusion using $\vv = \ve_i$.
To predict the functions of all proteins, we extend the linear system in Equation~\eqref{eq:ppr} to a matrix form:
\begin{equation}\label{eq:pprfunc}
  \left(\bmat{\mI_m & \vec{0} \\ \vec{0} & \mI_n} - \alpha
\overbar{  \bmat{\mG & \vec{0} \\ \mR_T^T & \mH}}
  \right) \bmat{\mX_G \\ \mX_H} = (1 - \alpha) \bmat{\mI_m \\ \vec{0}},
\end{equation}
where the bar over the block matrix still indicates the whole matrix is normalized to be column-stochastic.
%where $\tilde{\mG}$ and $\tilde{\mR}_T^T$ denote versions of the matrices $\mG$ and $\mR_T^T$ that have their columns scaled so that the matrix $\bmat{\tilde{\mG} \\ \tilde{\mR}_T^T }$ is column-stochastic.
%where $\tilde{\mG}$ and $\tilde{\mR}_T^T$ denote a joint column-wise normalization of $\mG$ and $\mR_T^T$, which is not identical to the individual column-normalization $\bar{\mG}$ and $\bar{\mR_T}^T$.
The lower block of the solution, $\mX_H$, is the output matrix of BirgRank for function prediction, and has the same dimensions as $\mR^T$. To further control the proportion of diffusion passing between the two layers of the bi-relational graph, we parameterize the model in Equation~\eqref{eq:pprfunc} as
\begin{equation}\label{eq:pprparameters}
\begin{split}
  &\left(\bmat{\mI_m & \vec{0} \\ \vec{0} & \mI_n} - \alpha
\overbar{ \bmat{\mu\mG & \vec{0} \\ (1-\mu)\mR_T^T & \mH^*}}
  \right) \bmat{\mX_G \\ \mX_H} \\
  &= (1 - \alpha) \overbar{\bmat{\theta\mI_m \\ (1-\theta)\mR_T^T}},
\end{split}
\end{equation}
where $\mH^* = \lambda \mH + (1-\lambda)\mH^T$, and $\lambda$ controls the diffusion direction on $\mH$. Specifically, $\lambda = 0$ indicates that the diffusion flows down the hierarchy, and $1$ indicates flow up the hierarchy. The parameter $\mu \in (0,1)$ controls the proportion of the diffusion flowing within $\mG$, and $\theta \in (0,1)$ controls the weighted sources between the proteins and functional annotations in the right-hand side of Equation~\eqref{eq:pprparameters}.

\subsection{Extension to AptRank} \label{sec:aptrank}

In the traditional model of PageRank, which we use in BirgRank, the teleportation parameter $\alpha \in (0,1)$ can be thought of as controlling the rate of decay of the diffusion as it spreads from the nodes in the personalization vector $\vv$ to the rest of the graph. After $k$ steps the diffusion has decayed by a factor of $\alpha^k$, for $k = 1,\cdots,\infty$ (Equation~\eqref{eq:prseries}). There are a variety of other empirical weighting schemes \citep{baeza2006generalizing,constantine2009random,chung2007heat,zhu2014adaptive}, each with slightly different theoretical properties.

In this section, we seek to replace the standard, fixed diffusion coefficients $\alpha^k$ at each step with an adaptive parameter, denoted by $\gamma^{(k)}$, to optimize the predictive power of the Markov chain.
To do this we repeatedly split the training set of protein function annotations, $\mR_T$, into different subsets to use in fitting and validating the coefficients. We denote the matrix used for fitting by $\mR_F$, and the matrix used in validation by $\mR_V$. These matrices have the same dimensions as $\mR_T$ and consist of entries of $\mR_T$, i.e., $\mR_T = \mR_F + \mR_V$.

%To determine the adaptive coefficients, we first split the matrix $\mR_T$ into the matrices $\mR_T$ and $|into a matrix $\mR_F$ used for fitting, and $\mR_V$ used for validating the coefficients $\gamma^{(k)}$.
%
%The matrix $\mR_F$ is a subset of the matrix $\mR_T$ that we use for fitting, and the matrix $\mR_V$ contains the remainder of the non-zero entries of $\mR_T$. The matrix $\mR_V$ we use for self-evaluation as follows.

To determine the adaptive coefficients $\gamma^{(k)}$ so that they bias predictions toward the training data, we proceed as follows.
The AptRank method begins by computing terms in the following sequence:
%using the system in Equation~\eqref{eq:pprfunc}:
\begin{equation}\label{eq:iter}
  \mX^{(k)} = \bmat{\mX_G^{(k)} \\ \mX_H^{(k)}} = \overbar{\bmat{\mG & \mR_F^* \\ \mR_F^T & \mH^*}} ^k \mX^{(0)},
\end{equation}
where the bar over the block matrices still denotes column-stochastic normalization,
\begin{equation}\label{eq:x0}
  \mX^{(0)} = \bmat{\mX_G^{(0)} \\ \mX_H^{(0)}} = \bmat{\mI_m \\ \vec{0}},
\end{equation}
and
\[ \mR_F^* =
  \begin{cases}
    \vec{0}       & \quad \text{to use a one-way diffusion}\\
    \mR_F  & \quad \text{to use a two-way diffusion}\\
  \end{cases}.
\]
We denote AptRank using a one-way diffusion and a two-way diffusion as AptRank-1 and AptRank-2, respectively. These two variations can have significant differences in prediction performance when the underlying networks have different sparsities.

To compute the optimal set of coefficients $\gamma^{(k)}$ that best fits the validation set $\mR_V$, we solve the following constrained least squares model,
\begin{equation}\label{eq:opt}
\begin{aligned}
& \underset{\gamma}{\text{minimize}}
& &  \left\| \text{vec}(\mR_V^T) - \sum_{i=k}^K \gamma^{(k)}\text{vec}(\mX_H^{(k)}) \right\|_2^2\\
& \text{subject to}
& & \sum_{k=1}^K \gamma^{(k)} = 1, \\
&&& \gamma^{(k)} \geq 0,
\end{aligned}
\end{equation}
where $\text{vec}(\cdot)$ is a matrix-to-vector transformation that stacks the columns of the matrix into a single column vector.

The entire AptRank framework is summarized in Algorithm~\ref{alg:aptrank}. We perform this fitting-validating process multiple times, each time splitting $t\%$ of entries in $\mR_T$ into new matrices $\mR_F$ and $\mR_V$ by choosing entries from $\mR_T$ uniformly at random. Each such iteration generates a new set of coefficients $\gamma^{(k)}$, which we store. We call these iterations ``shuffles" because in essence they consist of shuffling the entries of $\mR_T$ into the two matrices $\mR_F$ and $\mR_V$. Again, we note that the annotations in each row (for each protein) of $\mR_F$ and $\mR_V$ do not share parental ontology terms.
The number of shuffles performed, denoted as $S$, is an input parameter; after the prescribed number of shuffles is completed, we compute the average ${\gamma^*}^{(k)}$ of the $\gamma^{(k)}$ across all shuffles, and use those averaged ${\gamma^*}^{(k)}$ to compute the final diffusion values $\mX_{\text{AptRank}}$. This prediction solution will be compared against the evaluation set $\mR_E$ (see Section~\ref{sec:results}).

%The AptRank framework is as follows:
%\begin{enumerate}
%  \item Choose entries of $\mR$ uniformly at random to split $\mR$ into $\mR_T$ for training and $\mR_V$ for evaluation;
%  \item Choose entries of $\mR_T$ uniformly at random to split $\mR_T$ into $\mR_F$ for fitting and $\mR_V$ for validation;
%  \item Iteratively compute and store $\mX_H^{(i)}$ where $i = 0, 1, \cdots, k$ using Equation~(\ref{eq:iter});
%  \item Solve the constrained least squares system in Equation~\eqref{eq:opt};
%  \item Repeat Step $2$ to $4$ multiple times, storing the coefficients $\gamma^{*(i)}$ from each iteration;
%  \item Set $\gamma^{*(i)}$ to be the average of the coefficients $\gamma^{(i)}$ from all iterations;
%  \item Compute $\mX_{\text{AptRank}} = \sum \gamma^{*(i)}\mX_H^{*(i)}$, where $\mX_H^{*(i)}$ is computed using Equation~(\ref{eq:iter}) with $\mR_T$ in place of $\mR_F$;
%  \item Output the matrix $\mX_{\text{AptRank}}$ for use in predictions.
%\end{enumerate}

\begin{algorithm}[!ht]
\DontPrintSemicolon
\SetAlgoLined
%\KwResult{$\mX_{\text{AptRank}}$}
\SetKwInOut{Input}{Input}
\SetKwInOut{Output}{Output}
\SetKwFunction{SplitR}{splitR}
\SetKwFunction{Vec}{vec}
\SetKwFunction{Qr}{qr}
\SetKwFunction{Median}{median}

\Input{$\mG$,$\mR_T$,$\mH^*$,$K$,$S$,$t$}
\Output{$\mX_{\text{AptRank}}$}
\BlankLine

\For{$s \leftarrow 1$ \KwTo $S$}{
    [$\mR_F$, $\mR_V$] $\leftarrow$ \SplitR{$\mR_T$,$t$}\;
    \tcp*[h]{Choose $t\%$ of nonzero entries in $\mR_T$ uniformly at random and split to $\mR_F$, and derive $\mR_V = \mR_T - \mR_F$.}\;
    Initialize $\mX^{(0)}$ using Equation~\eqref{eq:x0}\;
    \For{$k \leftarrow 1$ \KwTo $K$}{
        Compute $\mX^{(k)}$ using Equation~\eqref{eq:iter}\;
        $\mA[:,k] \leftarrow $ \Vec{$\mX_H^{(k)}$}\;
    }
    [$\mQ_A$,$\mR_A$] $\leftarrow$ \Qr{$\mA$} \tcp*[h]{QR decomposition}\;
    $\vb$ $\leftarrow $ \Vec{$\mR_V$}\;
    Solve $\MINone{\vec{\gamma}^{(s)}}{\normof{\mQ_A^T \vb - \mR_A \vec{\gamma}^{(s)}}_2^2}{\mathlarger{\sum}_k \gamma_k^{(s)} = 1, \gamma_k^{(s)} \geq 0}$\;
    \tcp*[h]{Equivalently as Equation~\eqref{eq:opt}.}\;
}
$\vec{\gamma}^*$ $\leftarrow$ \Median{$\vec{\gamma}^{(s)}$}\;
\tcp*[h]{Take the median over all $s = 1$ to $S$ for each $k$.}\;
$\bmat{\mX_G^* \\ \mX_H^*} \leftarrow \mathlarger{\mathlarger{\sum}}_{k=1}^K \gamma^*_k \overbar{\bmat{\mG & \mR_T^* \\ \mR_T^T & \mH^*}}^k \bmat{\mI_m \\ \vec{0}}
$\;
%\colst{\bmat{\mG & \mR_T \\ \mR^T_n & \mH^*}}^k \mX^{(0)}
Output $\mX_{\text{AptRank}} \leftarrow \mX_H^*$ for use in prediction.
\caption{\label{alg:aptrank}AptRank}
\end{algorithm}

%%%%%%%%%%%%%%%%%%%%%%%%%%%%%%%%%%%%%%%%%%%%%%%%%%%%%%%%%

\subsection{Connection with Other Methods}
To investigate the similarities and differences of our methods and the other four previous methods used for evaluation, we perform a theoretical analysis and comparison here, and summarize the features of each method in Table~\ref{tab:summet}.
\begin{table*}[t]
		\centering
		\caption{Summary of the Six Methods \label{tab:summet}}
   \resizebox{1.0\textwidth}{!}{%
		\begin{tabular}{lccccccl}
		\hline
		\multicolumn{1}{c}{\textbf{\begin{tabular}[c]{@{}c@{}}Method\\ Name\end{tabular}}} & \textbf{\begin{tabular}[c]{@{}c@{}}Method\\ Type\end{tabular}} & \textbf{\begin{tabular}[c]{@{}c@{}}Functional\\ Hierarchy\end{tabular}} & \textbf{\begin{tabular}[c]{@{}c@{}}Bi-relational\\ Graph\end{tabular}} & \textbf{\begin{tabular}[c]{@{}c@{}}Negative\\ Samples\end{tabular}} & \textbf{\begin{tabular}[c]{@{}c@{}}Random\\ Walk\end{tabular}} & \textbf{\begin{tabular}[c]{@{}c@{}}Stationary\\ PageRank\end{tabular}} & \multicolumn{1}{c}{\textbf{Reference}} \\ \hline
		GeneMANIA-SW & \begin{tabular}[c]{@{}c@{}}kernel integration\\ \& classification\end{tabular} &  &  & \checkmark & \checkmark & \checkmark & \begin{tabular}[c]{@{}l@{}}\cite{genemania}\\ \cite{sw}\end{tabular} \\
		TMC & diffusion &  & \checkmark &  & \checkmark & \checkmark & \cite{yu2013protein} \\
		ProteinRank & regression &  &  &  & \checkmark & \checkmark & \cite{freschi2007protein} \\
		DCA-clusDCA & \begin{tabular}[c]{@{}c@{}}diffusion \&\\ decomposition\end{tabular} & \checkmark &  & \checkmark & \checkmark & \checkmark & \begin{tabular}[c]{@{}l@{}}\cite{dca}\\ \cite{wang2015exploiting}\end{tabular} \\
		BirgRank & diffusion & \checkmark & \checkmark &  & \checkmark & \checkmark & this study \\
		AptRank & diffusion & \checkmark & \checkmark &  & \checkmark &  & this study \\ \hline
		\end{tabular}%
}% end resizebox
\end{table*}

The linear system of BirgRank in Equation~\eqref{eq:pprfunc} can be expanded into
\begin{equation}\label{eq:twoeq}
  \begin{aligned}
  \begin{cases}
    (\mI - \alpha \tilde{\mG})\mX_G = (1-\alpha)\mI\\
    \alpha \tilde{\mR_T}^T \mX_G = (\mI - \alpha \tilde{\mH} )\mX_H,
  \end{cases}
  \end{aligned}
\end{equation}
where $\tilde{\mG}$, $\tilde{\mR_T}$, and $\tilde{\mH} = \overbar{\mH}$ denote the submatrices of the column-stochastic matrix in Equation~\eqref{eq:pprfunc}.
%$\colst{\bmat{\mG & \vec{0} \\ \mR_T^T & \mH}}$ corresponding to the positions of $\mG$ and $\mR_T^T$.
By solving Equations~\eqref{eq:twoeq} for $\mX_H$, we get
\begin{equation}\label{eq:split}
  \mX_H = \alpha(1-\alpha)(\mI - \alpha\overbar{\mH})^{-1} \tilde{\mR_T}^T (\mI - \alpha\tilde{\mG})^{-1}.
\end{equation}

In contrast, ProteinRank \citep{freschi2007protein} uses only the protein-protein association network $\mG$ as a one-layer network model --- and does not directly take into consideration the functional hierarchy $\mH$ --- and then computes PageRank using $\mR_T$ as the personalization vectors (matrix). ProteinRank constructs a regression model and solves the linear system
\begin{equation}\label{eq:proteinrank}
  \mX_{\text{ProteinRank}} = (1-\alpha) (\mI - \alpha \overbar{\mG})^{-1} \mR_T,
\end{equation}
which can cause poor prediction quality due to the assumption of independence between functions (see Section~\ref{sec:results}). Our method BirgRank is closely related to ProteinRank:
if we plug $\overbar{\mH} = \mI$ into Equation~\eqref{eq:split}, then the resulting BirgRank solution differs from the ProteinRank solution (Equation~\eqref{eq:proteinrank}) only by a scalar coefficient and a slightly different normalization of $\mG$.

Similar to ProteinRank, GeneMANIA \citep{genemania} models protein function prediction as a multiclass-multilabel classification problem by integrating multiple heterogeneous network datasets and then using the Label Propagation algorithm \citep{zhou2004learning} as
\begin{equation}\label{eq:genemania}
  \mX_{\text{GeneMANIA}} = (\mI - \mL)^{-1} \mR^{*T},
\end{equation}
where $\mL = \mD - \mW$ is the Laplacian matrix, $\mW$ is a weighted sum of multiple kernel matrices from heterogeneous network data sets, and $\mD$ is a diagonal matrix with $D_{ii} = \sum_j W_{ij}$. Additionally, GeneMANIA extends the binary matrix $\mR_T^T$ to $\mR^{*T}$ by introducing negative samples in which $R_{i,j}^* = -1$ if protein $i$ is known not to have function $j$. The developers of GeneMANIA further accelerated their algorithm by introducing Simultaneous Weights (hereafter GeneMANIA-SW) \citep{sw}.

\citet{yu2013protein} proposed the Transductive Multilabel Classifier (TMC) by directly applying a Bi-relational graph model used in image annotation \citep{wang2011image} to protein function prediction, without consideration of the functional hierarchy. Instead, they use the cosine similarity of functional annotations to construct a function-function similarity matrix to replace $\mH$. The key difference between TMC and BirgRank is that TMC allows information to diffuse from functional terms to proteins, but not proteins to functional terms, as in BirgRank. Mathematically, the transition matrix of PageRank used in TMC is
\begin{equation}\label{eq:tmc}
  \overbar{\mA_{\text{TMC}}} = \overbar{\bmat{\mW_G & \mW_R \\ \vec{0} & \mW_F}},
\end{equation}
where the matrix $\mW_F$ is the degree-weighted function-function cosine similarity, i.e., $\cos(\mR_T^T,\mR_T)$, $\mW_G$ is a degree-weighed graph kernel of protein-protein association network, and $\mW_R$ is a normalized function profile derived from $\mR_T$. The developers of TMC suggest further flattening the functional hierarchy by using a random walk with restart approach \citep{yu2015down}. But this method, called dRW, does not use a bi-relational graph model, and was tested only using a very small data set \citep{yu2015down}.

\citet{wang2015exploiting} proposed clusDCA by extending their original Diffusion Component Analysis (DCA) method \citep{dca}. The clusDCA algorithm first uses PageRank to smooth both of the graphs, denoted as $\mG$ and $\mH$ in this study. Next, it computes Singular Value Decomposition (SVD) for the two smoothed matrices for low-rank matrix approximations. Finally, it attempts to find the optimal projection between the two low-rank matrices.

%%
%%  RESULTS SECTION
%%
\section{Results}\label{sec:results}

\subsection{Experimental Setup}
We present a comprehensive evaluation of the six methods using the three benchmark datasets from yeast, human and fly that can be downloaded from the GeneMANIA-SW website \citep{swdataweb}.
All three datasets were collected by the developers of GeneMANIA in 2010. We collected one more dataset for human proteins from public databases in March 2015 in order to test all the methods using up-to-date data with a larger size than those collected in 2010 (see Table~\ref{tab:dataset}). In this human dataset, denoted as human-2015, the network $\mG$ was downloaded from BioGRID \citep{biogrid}, and the annotations $\mR$ and the hierarchy $\mH$ from the Gene Ontology Consortium \citep{gene2015gene}.
The number of direct GO (Table~\ref{tab:dataset}, $3$rd column) indicates the number of annotations of individual proteins directly downloaded from the Gene Ontology Annotation (GOA) database. This does not reflect the implied inclusion of parental terms (see the total number of terms in Table~\ref{tab:dataset}, $4$th column for comparison). The multiple kernels (Table~\ref{tab:dataset}, $5$th column) from heterogeneous molecular data were directly downloaded from the GeneMANIA-SW website \citep{swdataweb}, and combined into a single network (i.e., $\mG$) with the weights provided in the datasets.

\begin{table}[ht]
\centering
\caption{Statistics of datasets}
\label{tab:dataset}
\begin{tabular}{lcccc}
\hline
\multicolumn{1}{c}{\textbf{\begin{tabular}[c]{@{}c@{}}Data\\ Set\end{tabular}}} & \textbf{\begin{tabular}[c]{@{}c@{}}No. of\\ proteins\end{tabular}} & \textbf{\begin{tabular}[c]{@{}c@{}}No. of\\ direct GO\end{tabular}} & \textbf{\begin{tabular}[c]{@{}c@{}}No. of\\ all GO\end{tabular}} & \textbf{\begin{tabular}[c]{@{}c@{}}No. of\\ kernels\end{tabular}} \\ \hline
\textbf{Yeast} & 3904 & 1188 & 1695 & 44 \\
\textbf{Human-2010} & 13281 & 1952 & 2919 & 8 \\
\textbf{Fly} & 13562 & 2195 & 2919 & 38 \\
\textbf{Human-2015} & 14515 & 11519 & 27106 & 1 \\ \hline
\end{tabular}
\end{table}

\begin{figure}[!ht]
        \centering
        \includegraphics[width = 0.75\textwidth]{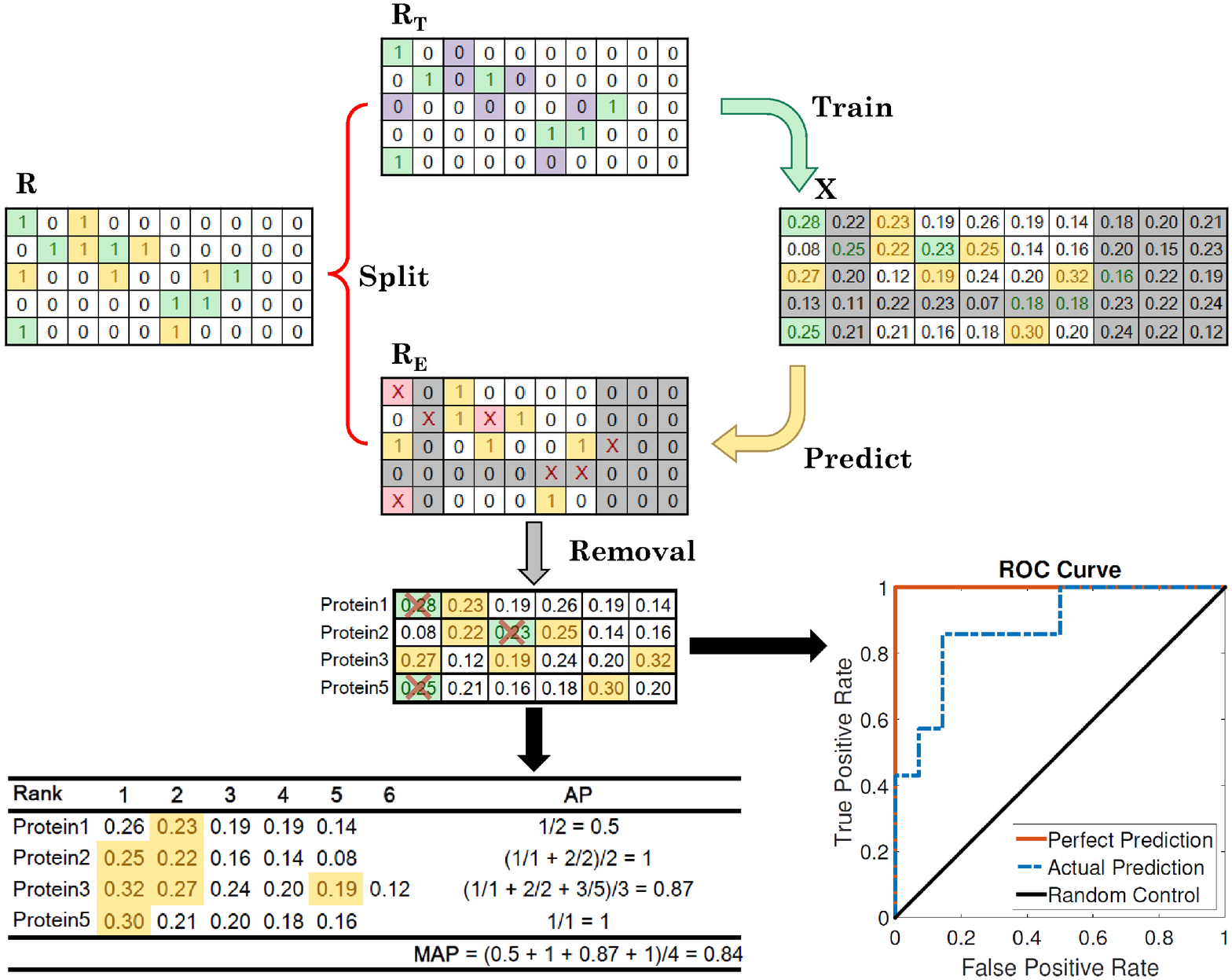}\\
        \caption{Missing Function Prediction Strategy. Split the given annotations $\mR$ by putting $50\%$ into the training set $\mR_T$ and $50\%$ into the evaluation set $\mR_E$. Then compare the predictions against $\mR_E$ and evaluate the performance of each method using AUROC and MAP.}
        \label{fig:split}
    \end{figure}

To evaluate the quality of each method in protein function prediction, we conducted cross validation using three different strategies to split the given functional annotation data $\mR$ into $\mR_T$ used for training and $\mR_E$ used for evaluation (see Section~\ref{sec:perf}). The three strategies are:
\begin{enumerate}
 \item \hspace*{1em} missing function prediction
 \item \hspace*{1em} \emph{de novo} function prediction
 \item \hspace*{1em} guided function prediction.
\end{enumerate}
All three validation strategies ensure that the matrices $\mR$, $\mR_T$ and $\mR_E$ have the same dimensions, and $\mR = \mR_T + \mR_E$. To measure the prediction quality of each method, we use two evaluation metrics: AUROC (Area Under the Receiver Operating Characteristic curve) which is widely used in protein function prediction, and MAP (Mean Average Precision) which is widely used in information retrieval (Figure~\ref{fig:split}). The key advantage of MAP is that MAP does not take true negatives into account, and is thus a more informative metric than AUROC when negative samples outnumber positive samples. This is true in our case since in the human-2015 dataset, for example, we attempt to predict $45$ functions on average from $11,519$ possible annotations (feature space, see Table~\ref{tab:dataset}).

We determined parameter settings as follows. For the four methods other than our BirgRank and AptRank, we mostly used the default settings specified in the corresponding literature.
%In terms of parameter setting, we mainly used default setting for the four compared methods, as specified in the corresponding literature.
We only tuned the reduced dimensionality $d$ in clusDCA to be $500$, rather than the parameter setting $2,500$ specified by the authors \citep{wang2015exploiting}, since this parameter is a key factor in time complexity of clusDCA. Empirically, we found that clusDCA is the most time-consuming method as shown in Table~\ref{tab:ctime}, and a large $d$ value dramatically increases running time. For the parameters in BirgRank, we set $\lambda = 0.5$ in determining $\mH^*$, to allow equal diffusion upward and downward the hierarchy. For the other three parameters $\alpha$, $\theta$, and $\mu$ in BirgRank (See Equation~\eqref{eq:pprparameters}), we observed that different settings of these three parameters did not yield significant differences in performance, and found that a value of $0.5$ empirically achieved good results. For the parameters in AptRank, we set the total iteration number $K$ to be $8$, the splitting parameter $t$ to be $50\%$, and the number of shuffles $S$ to be $5$. These setting may vary depending on the validation strategies and the data sizes, which we discuss in Section~\ref{sec:perf}.

%%%%%%%%%%%%%%%%%%%%%%%%%%%%%%%%%%%%%%%%%%%%%%%%%%

\subsection{Comparison of Prediction Performances}\label{sec:perf}
\subsubsection{Missing Function Prediction}\label{sec:miss}

We first conducted a numerical experiment to evaluate the ability of the six methods in predicting missing protein functions as follows. We uniformly select a certain percentage of non-zero entries in $\mR$ at random, move them to a matrix $\mR_T$ for training, and let $\mR_E = \mR - \mR_T$ be the evaluation set. Figure~\ref{fig:split} illustrates how to split matrix $\mR$ with $14$ entries into $\mR_T$ and $\mR_E$ when the splitting percentage is specified as $50\%$. We carried out this random sampling with replacement $5$ times for each specified splitting percentage. This is not a circular cross validation since it does not guarantee that each functional annotation is tested once and only once. This strategy aims to test whether the methods can restore incomplete functional annotations for each protein and is unbiased with respect to how many annotations each protein has.

    \begin{figure*}[!ht]
        \centering
        \includegraphics[width = 0.9\textwidth]{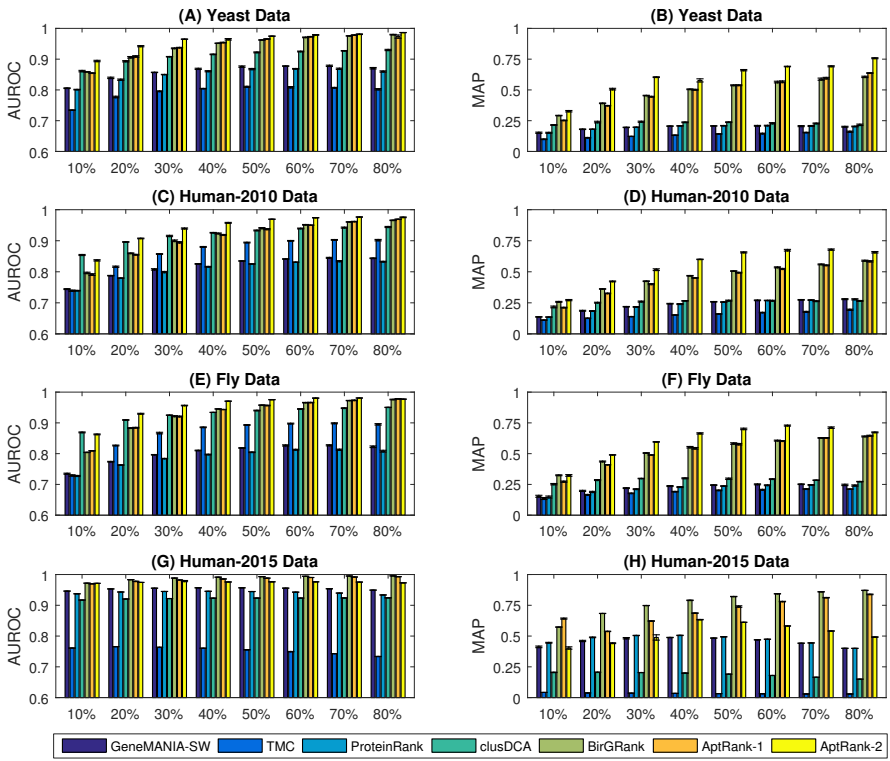}\\
        \caption{Missing function prediction. The \emph{x}-axis represents the percentages of data used in training. The error mark on top of each bar indicates the standard deviation of AUROCs or MAPs over 5 repetitions of each experiment.}
        \label{fig:miss_perf}
    \end{figure*}

We start with $10\%$ split for training and increase by increments of $10\%$ up to $80\%$ (Figure~\ref{fig:miss_perf}). Generally, the resulting AUROCs and MAPs of the six methods show that both BirgRank and AptRank outperform the other four previous methods in all $8$ groups of experiments with different amounts of training data. In the $10\%$ group of human-2010 and fly datasets, clusDCA slightly outperforms our methods in AUROC, but its MAP is lower than those of our methods (Figure~\ref{fig:miss_perf} (C) and (E)). When more data are given for training, our methods outperform the other four methods in terms of MAP with approximately 2- to 3-fold improvement.

To investigate the effect of the GO functional hierarchy in prediction, we compare the performance of non-hierarchy-integrated methods (GeneMANIA-SW, TMC and ProteinRank) with hierarchy-integrated methods (clusDCA, BirgRank and AptRank). We find that the integration of the functional hierarchy clearly improves the prediction accuracy (Figure~\ref{fig:miss_perf}). Furthermore, our methods, for the most part, perform better than clusDCA, which suggests that using a bi-relational graph framework (Figure\ref{fig:example}) to integrate the hierarchy is better than seeking for projection between the protein network and the functional hierarchy.

Comparing the performances of BirgRank and AptRank, we find that the performance of the algorithms differs as the network sparsity varies (Figure~\ref{fig:miss_perf} (B), (D), (F) vs. (H)). The three benchmark datasets are smaller and denser than Human-2015 dataset due to the integration of multiple kernels (Table~\ref{tab:dataset}). We can see that AptRank with a two-way diffusion performs better on the dense network, while BirgRank is better on the sparse network. This could be because a dense network restricts network diffusion within a local region of the source node, and two-way diffusion forms a feedback loop that enhances the contributions of the annotations within local regions. However, the two-way diffusion spreads out of this local region in a sparse network and provides irrelevant feedback to the source node.

In addition, we find that GeneMANIA-SW and ProteinRank achieve similar performance in both AUROC and MAP. The key difference between these two models is that GeneMANIA-SW requires negative samples in its classification framework. This demonstrates that negative samples have a very limited contribution to the performance of GeneMANIA-SW on these datasets.
%Indeed, it is difficult to confirm that one protein does not have a function.
This could be in part because it can be difficult to confirm that a protein does not have a function.

Lastly, we find that BirgRank outperforms TMC. Theoretically, the models of TMC and BirgRank are quite similar, differing mainly in how the two methods direct the diffusion between the two network layers, $\mG$ and $\mH$. BirgRank diffuses information from $\mG$ to $\mH$, while TMC does the reverse. Our results support the idea that diffusion from proteins to functional terms is the more useful direction in the context of protein function prediction.

%%%%%%%%%%%%%%%%%%%%%%%%%%%%%%%%%%%%%%%%%%%%%%%%%%%%

\subsubsection{\textit{De novo} Function Prediction}\label{sec:denovo}

\begin{figure*}[!ht]
        \centering
        \includegraphics[width = 0.9\textwidth]{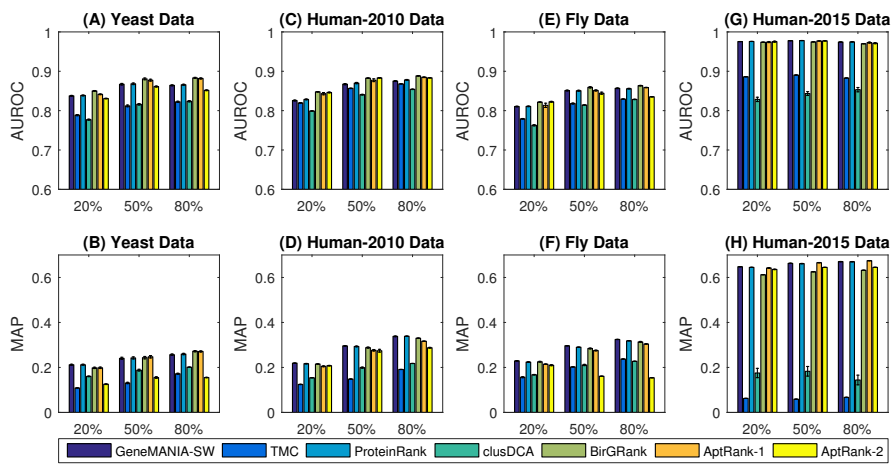}\\
        \caption{\textit{De novo} function prediction. The \emph{x}-axis represents the percentages of data used in training. The error mark on top of each bar indicates the standard deviation of AUROCs or MAPs over 3 repetitions of each experiment.}
        \label{fig:denovo_perf}
    \end{figure*}

To investigate whether the six methods can accurately predict the functions of one protein without any annotation for training, we design a \textit{de novo} circular cross validation as follows. Uniformly partition a certain percentage, denoted as $c$, of proteins into $b$ groups at random. Letting $[v]$ denote the nearest-integer operation
%is operation of rounding a real number $v$ to its nearest integer, and
we can calculate
\[ b =
  \begin{cases}
    \left[1/c\right]       & \quad \text{if } 0 < c \leq 0.5\\
    \left[1/(1-c)\right]  & \quad \text{if } 0.5 < c \leq 1\\
  \end{cases}.
\]
In practice, we set $c$ as $20\%$, $50\%$ and $80\%$ as shown in the $x$-axis of  Figure~\ref{fig:denovo_perf}. When $c = 80\%$, it is equivalent to a conventional five-fold cross validation with $80\%$ of proteins as the training set and the complementary $20\%$ as the evaluation set. On the contrary, $c = 20\%$ means we only use $20\%$ of proteins for training and evaluate the prediction performance by the complementary $80\%$. Lastly, $c = 50\%$ is equivalent to a two-fold cross validation. Normally, three-fold cross validation ($c = 66.7\%$) is used in the four reference methods. Here, our cross validation design is aimed to explore the potential predictive power of all of the methods with a more stringent criterion.

As shown in Figure~\ref{fig:denovo_perf}, our methods generally perform no worse than the four reference methods. Interestingly, GeneMANIA has nearly the same performance as ProteinRank in both AUROC and MAP metrics, which occurs in our missing function prediction experiment as well (Figure~\ref{fig:miss_perf}).
Furthermore, they both perform better than the other two reference methods, TMC and clusDCA. Our methods perform slightly better than GeneMANIA and ProteinRank in AUROC, but do slightly worse in MAP.
This leads us to conclude that (1) a classification model that includes negative samples (GeneMANIA) is little different from a diffusion model (ProteinRank) in \textit{de novo} function prediction; and (2) integrating the GO hierarchy (BirgRank and AptRank) cannot significantly improve the accuracy in function prediction for newly found proteins without known functional information.

%%%%%%%%%%%%%%%%%%%%%%%%%%%%%%%%%%%%%%%%%%

\subsubsection{Guided Function Prediction}\label{sec:guided}

To examine the extent to which our methods benefit from limited known annotations of tested proteins, we devise a validation strategy called guided function prediction which is a hybrid of the missing function prediction (Section~\ref{sec:miss}) and the \textit{de novo} prediction (Section~\ref{sec:denovo}) strategies. In this validation, the strategy of partitioning training and evaluation sets is identical to that used in \textit{de novo} prediction except that it gives \emph{one} functional annotation as guidance for each evaluated protein that has more than one annotation. The proteins in the evaluation set with only one or no annotation are not taken into account.

    \begin{figure*}[!ht]
        \centering
        \includegraphics[width = 0.9\textwidth]{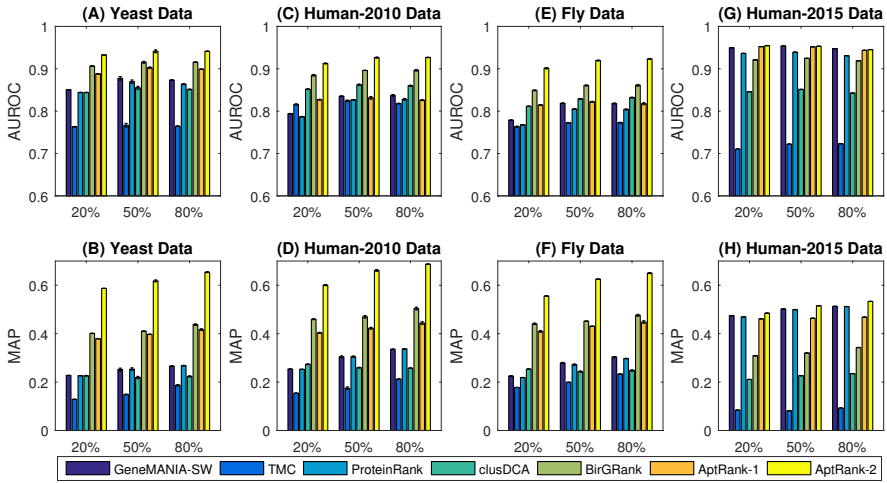}\\
        \caption{Guided function prediction. The \emph{x}-axis represents the percentages of data used in training. The error mark on top of each bar indicates the standard deviation of AUROCs or MAPs over 3 repetitions of each experiment.}
        \label{fig:guided_perf}
    \end{figure*}

We can see in Figure~\ref{fig:guided_perf} that in the evaluations using the three benchmark datasets with dense network data, our methods, especially AptRank-2, can take full advantages of the single given annotation to improve prediction performance by approximately 2-fold in AUROC and 3-fold in MAP, compared to the other four methods. In the sparse network data (Human-2015), we find that the given annotations worsen the performances of all the methods (Figure~\ref{fig:denovo_perf} (G,H) vs. Figure~\ref{fig:guided_perf} (G,H)). We conclude that sparse network datasets may cause underfitting of our model training, and reducing the model complexity can alleviate this problem, e.g., setting a small $\alpha$ in BirgRank or a small $K$ in AptRank. On the contrary, we also find that in some experiments, the more data we provide for training, the worse the testing accuracy is (e.g., AptRank-2 in Figure~\ref{fig:denovo_perf}(F)). In these cases, \citet{verleyen2015positive} proposed using sampling of the training data to overcome this overfitting.
%[for some unknown reason...]

Finally, all three validations show that AUROC is always higher than MAP in the evaluation of the same prediction result. This suggests that MAP is a better metric when the number of negative samples is much larger than the number of positive samples, as is the case in protein function prediction.

%%
%%  RESULTS SECTION
%%

\subsection{Analysis of Adaptive Coefficients}

\begin{table}[!ht]
\centering
\caption{Medians of $\gamma$ in Prediction of Yeast and Human-2015 Datasets}
\label{tab:gamma}
\begin{tabular}{cccccccccc}
\hline
\multirow{2}{*}{\textbf{\begin{tabular}[c]{@{}c@{}}Data\\ Set\end{tabular}}} & \multirow{2}{*}{\textbf{\begin{tabular}[c]{@{}c@{}}Training\\ (\%)\end{tabular}}} & \multicolumn{8}{c}{\textbf{Markov chain iteration}} \\
 &  & \textbf{1st} & \textbf{2nd} & \textbf{3rd} & \textbf{4th} & \textbf{5th} & \textbf{6th} & \textbf{7th} & \textbf{8th} \\ \hline
\multirow{8}{*}{\textbf{Yeast}} & \textbf{10\%} & 0 & 0 & 0 & 0 & 0 & 0 & 0.08 & 0.92 \\
 & \textbf{20\%} & 0 & 0.11 & 0 & 0 & 0 & 0 & 0.23 & 0.66 \\
 & \textbf{30\%} & 0 & 0.34 & 0 & 0.08 & 0 & 0 & 0.58 & 0 \\
 & \textbf{40\%} & 0 & 0 & 0 & 0 & 0 & 0 & 1 & 0 \\
 & \textbf{50\%} & 0 & 0 & 0 & 0 & 0.84 & 0 & 0.16 & 0 \\
 & \textbf{60\%} & 0 & 0 & 0 & 0 & 1 & 0 & 0 & 0 \\
 & \textbf{70\%} & 0 & 0 & 0.09 & 0 & 0.91 & 0 & 0 & 0 \\
 & \textbf{80\%} & 0 & 0 & 0.64 & 0 & 0.36 & 0 & 0 & 0 \\ \hline
\multirow{8}{*}{\textbf{\begin{tabular}[c]{@{}c@{}}Human\\ 2015\end{tabular}}} & \textbf{10\%} & 0 & 0.20 & 0 & 0 & 0 & 0 & 0.31 & 0.49 \\
 & \textbf{20\%} & 0 & 0.65 & 0 & 0 & 0 & 0 & 0.11 & 0.24 \\
 & \textbf{30\%} & 0 & 1 & 0 & 0 & 0 & 0 & 0 & 0 \\
 & \textbf{40\%} & 0 & 1 & 0 & 0 & 0 & 0 & 0 & 0 \\
 & \textbf{50\%} & 0 & 1 & 0 & 0 & 0 & 0 & 0 & 0 \\
 & \textbf{60\%} & 0 & 1 & 0 & 0 & 0 & 0 & 0 & 0 \\
 & \textbf{70\%} & 0 & 1 & 0 & 0 & 0 & 0 & 0 & 0 \\
 & \textbf{80\%} & 0 & 1 & 0 & 0 & 0 & 0 & 0 & 0 \\ \hline
\end{tabular}
\end{table}

The adaptive coefficients of AptRank ($\gamma$) are the unique feature that differs from traditional PageRank. To investigate their behaviors in prediction, we list the medians of $\gamma$ over the different shuffles in the prediction of yeast and human-2015 datasets in Table~\ref{tab:gamma}. We can see that there are three main features of $\gamma$'s behaviors,
\begin{enumerate}
  \item~$\gamma^{(1)}$ is always zero, since the information diffusing within $\mG$, from proteins at the first step, has not yet reached the hierarchy;
  \item~as shown in the yeast dataset, the distribution of $\gamma$ is not uniform, but concentrates on specific terms of Markov chains, which demonstrates that AptRank can adaptively select the most predictive terms rather than weighting all terms with power-decays like traditional PageRank; and
  \item~in comparison of $\gamma$ in yeast and human-2015 datasets, we find that AptRank mostly selects the 2nd term in the human-2015 dataset, but a few more terms in the yeast dataset, which is due to the different network densities of the two datasets. The yeast dataset is smaller but denser, since it integrates $44$ different kernels into $\mG$; the human-2015 dataset is larger but sparser, and all the entries in the raw human-2015 dataset are binary. This implies that for a sparse dataset, our AptRank might be equivalent to neighbor-voting methods.
\end{enumerate}

%%%%%%%%%%%%%%%%%%%%%%%%%%%%%%%%%%%%%%%%%%%%%%%%%

\subsection{Comparison of Runtimes}
The average computational time of the six methods compared in this study are shown in Figure~\ref{tab:ctime}. In this comparison, the computational time is recorded for the prediction using the largest dataset, human-2015. We can clearly see AptRank requires the third longest computational time, likely because it involves many dense matrix operations. The SVD computations required in clusDCA are likely responsible for clusDCA having the longest running time. Without a parallel implementation of SVD, clusDCA might be impractical unless we sacrifice prediction accuracy by using a small $d$ value. GeneMANIA-SW is the second most computationally expansive method, since it computes the prediction scores function by function. This is extremely expensive when the number of functions is large, even though we only used direct GO terms in GeneMANIA-SW. BirgRank and TMC both use bi-relational graphs, and take only several minutes to solve the PageRank linear system. ProteinRank has the most simple model, and it takes the shortest time, since it needs only to solve a PageRank linear system with approximately half the dimension of the systems involved in BirgRank and TMC.

\begin{table}[!ht]
\centering
\caption{Runtimes of the Six Methods in Minutes (Human-2015 Dataset)*}
\label{tab:ctime}
\begin{tabular}{lrrrrrr}
\hline
\multicolumn{1}{c}{\multirow{2}{*}{\textbf{Methods}}} & \multicolumn{6}{c}{\textbf{Training Data Proportion}} \\
\multicolumn{1}{c}{} & \multicolumn{1}{c}{\textbf{10\%}} & \multicolumn{1}{c}{\textbf{20\%}} & \multicolumn{1}{c}{\textbf{40\%}} & \multicolumn{1}{c}{\textbf{50\%}} & \multicolumn{1}{c}{\textbf{70\%}} & \multicolumn{1}{c}{\textbf{80\%}} \\ \hline
\textbf{GM-SW} & 252.52 & 214.47 & 232.02 & 231.65 & 225.54 & 234.56 \\
\textbf{TMC} & 6.71 & 7.10 & 7.52 & 7.58 & 7.37 & 7.12 \\
\textbf{ProteinRank} & 0.85 & 0.87 & 0.87 & 0.87 & 0.88 & 0.88 \\
\textbf{clusDCA} & 1054 & 1019 & 1072 & 1061 & 1025 & 1050 \\
\textbf{BirgRank} & 9.42 & 9.46 & 9.46 & 9.45 & 9.42 & 9.49 \\
\textbf{AptRank-1} & 51.79 & 53.48 & 55.82 & 55.28 & 57.85 & 58.69 \\ \hline
\end{tabular}
\begin{flushleft}
{\footnotesize  *The runtimes of $30\%$ and $60\%$ is not shown due to space limit. The AptRank-1 uses 12-core parallel computing for matrix multiplication.}
\end{flushleft}
\end{table}

\section{Conclusion}

In this paper we present two network-diffusion-based methods for protein function prediction. Our first method, BirgRank, uses PageRank on a bi-relational graph model that incorporates protein-protein and function-function networks. Our second method, AptRank, introduces an adaptive mechanism to the PageRank framework that computes an optimal set of weights for the first several steps of diffusion so as to maximize recovery of a subset of known function annotations. We show that both methods outperform the four existing state-of-the-art methods in almost all cases, and in particular, outperform those methods that do not incorporate information about the functional hierarchy. Our results also suggest that diffusion-based methods are still among the most competitive in network-based protein function predictions, compared to classification-based and decomposition-based methods.

Furthermore, our methods provide a theoretical framework in data integration, which may benefit multi-omics studies in complex diseases, or multi-species metabolic network modeling in microbiome studies. From a general view outside bioinformatics, our methods can be used to develop multi-class recommendation systems in social media with inter-dependent labels. For example, the protein-protein association network in this study can be viewed as similar to the professional social network between LinkedIn users, and the functional hierarchy can be seen as generalizing to an individual's skill set. Those skill sets are typically inter-dependent. For instance, a user with knowledge of Perl programming is likely to have bioinformatics expertise.

\section*{Acknowledgements}

The authors would like to thank Dr. Jesse Gillis at Cold Spring harbor Laboratory for his helpful comments, and Purdue Rosen Center for Advanced Computing (RCAC) for their technical supports.

\section*{Funding}

This research has been supported in part by NSF CAREER CCF-1149756, NSF CAREER CCF-093937, and DARPA SIMPLEX.

\bibliographystyle{natbib}

\bibliography{aptref}      % Bibliography file (usually '*.bib' )

\end{document}